\documentclass[twocolumn,showpacs,preprintnumbers,amsmath,amssymb,aps,dvips,prl]{revtex4}
\usepackage{amsmath}
\usepackage[dvips]{graphicx}
\usepackage{dcolumn}
\newcolumntype{d}[1]{D{.}{.}{#1}}
\usepackage{bigdelim}
\usepackage{bm}
\usepackage{array}
\usepackage{longtable}
\usepackage{lscape}
\usepackage{float}
\usepackage{rotating}
\usepackage{multirow}
\usepackage{setspace}

\makeatletter

\newcommand{\Rmnum}[1]{\expandafter\@slowromancap\romannumeral #1@}
\makeatother

\begin{document}

\preprint{APS/123-QED}

\title{Entanglement Perturbation Theory for Infinite Quasi-1D Quantum Systems}
\author{Lihua Wang$^{1}$}
\email{wanglihua94@tsinghua.org.cn}
\author{Sung Gong Chung$^{2,3}$}

\affiliation{$^1$
Computational Condensed-Matter Physics Laboratory, RIKEN, Wako, Saitama 351-0198, Japan}
\affiliation{$^2$
Department of Physics and Nanotechnology Research and Computation Center, Western Michigan University, Kalamazoo, Michigan 49008, USA  
}%

\affiliation{$^3$
Asia Pacific Center for Theoretical Physics, Pohang, Gyeonbuk 790-784, South Korea}

\date{\today}

\begin{abstract} 
We develop Entanglement Perturbation Theory (EPT) for infinite Quasi-1D quantum systems. 
The spin $1/2$ Heisenberg chain with ferromagnetic nearest neighbor (NN) and antiferromagnetic
next nearest neighbor 
(NNN) interactions with an easy-plane anisotropy 
is studied as a prototypical system. The obtained accurate phase diagram is compared with a recent prediction [Phys.Rev.B, \textbf{81},094430(2010)] that dimer and N\'{e}el orders appear alternately as the XXZ anisotropy $\Delta$ approaches the isotropic limit $\Delta=1$. The first and second transitions (across dimer, N\'{e}el, and dimer phases) are detected with improved accuracy at $\Delta\approx 0.722$ and $0.930$. The third transition (from dimer to N\'{e}el phases), previously predicted to be at $\Delta\approx 0.98$, is not detected at this $\Delta$ in our method, raising the possibility that the second N\'{e}el phase is absent. 
\end{abstract}
\maketitle

\begin{figure}
\begin{center}
\includegraphics[height=1.2in, width=1.8in]{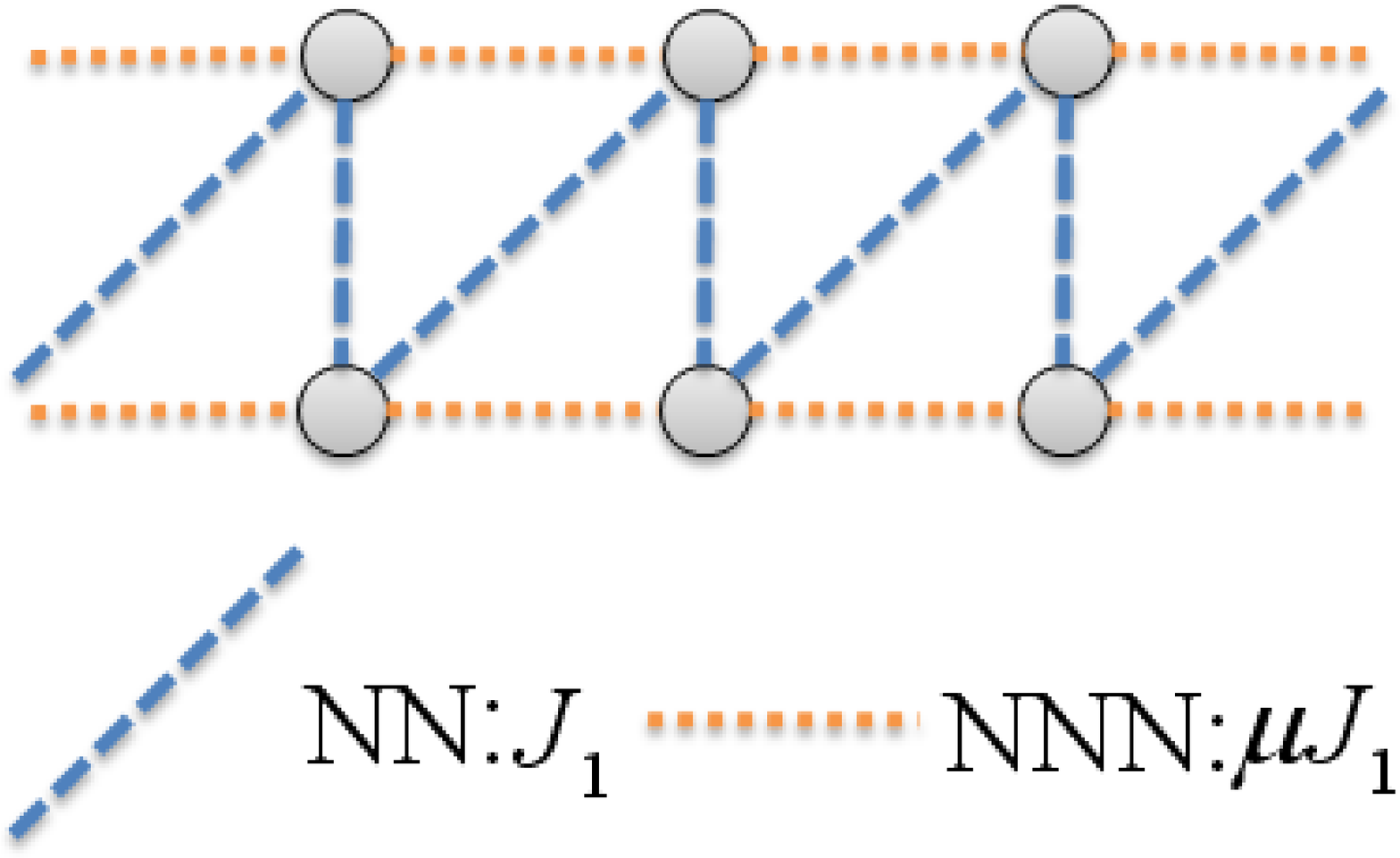}
\caption{\label{fig:deformation}(color online) A zigzag chain suitable for EPT. Circles refer 
to spins. Long dashed lines (blue) indicate the NN coupling $J_1$, while short ones (orange) 
the NNN coupling $J_2=\mu J_1$.}
\end{center}
\end{figure}

In a recent article, Furukawa et al. \cite{Furukawa} reported exotic N\'{e}el and dimer orders in a 
spin $1/2$ Heisenberg chain described by the Hamiltonian,
\begin{align}
\label{EQ:H}
H=\sum_{\delta=1}^{2}\sum_{i=1}^{L}J_{\delta}\left(S_i^xS_{i+\delta}^x+S_i^yS_{i+\delta}^y+\Delta S_i^zS_{i+\delta}^z\right)
\end{align}
characterized by a ferromagnetic NN exchange coupling $J_1<0$ and an antiferromagnetic 
NNN exchange coupling $J_2>0$, with an easy-plane anisotropy $\Delta<1$.
Let us describe the effect of the NNN exchange by $\mu\equiv J_2/J_1$. 
Fig.\ref{fig:deformation} shows an equivalent coupled 2 spin-chains (zigzag chain) suited for EPT.
The theoretical tools employed are the bosonization-Sine-Gordon (SG) theory \cite{Haldane}, 
the RG (renormalization group)-level spectroscopy analysis augmented by exact diagonalization
\cite{Okamoto,Nomura}, and numerical RG analysis based on iTEBD 
(infinite time evolving block decimation) \cite{Vidal}, a variance of DMRG (density matrix renormalization
group) method \cite{White1} for infinite 1D systems.
The key finding is that the N\'{e}el order and dimer order alternate numerous times 
when approaching the ferromagnetic transition point $(J_2/|J_1|,\Delta)=(1/4,1)$ along the Lifshitz line 
which separates the character of the spin-spin correlation in the xy-plane from 
commensurate to incommensurate. The reported N\'{e}el order is indeed counterintuitive
since neither ferromagnetic NN nor antiferromagnetic NNN exchange couplings favor antiferromagnetism.

When $Js$ are both positive, the ground state phase diagram of the model is well 
understood \cite{Majumdar,Haldane,White}. Particularly, for small $\mu$ and with easy-plane 
anisotropy $\Delta<1$, it is a gapless Tomonaga-Luttinger liquid (TLL). 
The gapless TLL phase has instabilities toward ordered states \cite{Majumdar}. 
N\'{e}el order will be induced by easy-axis anisotropy $\Delta>1$, whereas dimer order 
appears when $\mu$ is larger than some critical $\Delta$-related value. 
For instance, transition takes place when $\mu\approx 0.24$ at $\Delta=1$. 
The quantum phase transition to these ordered phases can be understood 
within a bosonization-SG theory \cite{Haldane}.

Now in the case $J_1<0$ of our interest, 
the bosonization-SG theory of the TLL phase at $J_2=0$ is known exact, and the "infinite" alternation of
energies of the N\'{e}el and dimer states was observed in the parameter space $0<\Delta<1$, and 
based on a perturbative treatment, from the Gaussian Hamiltonian, of the cosine term in the SG theory, 
the authors of \cite{Furukawa, Furukawa1} claimed that the same should occur deep inside the frustrated 
regime $|\mu|\sim1/4$, and thus
the realization of the true alternating appearance of the N\'{e}el and dimer phases. This prediction
was first checked by the level spectroscopy \cite{Okamoto,Nomura}, showing 5 times of phase alternation 
between the two phases. This level spectroscopy analysis was, however, still based on
the effective SG theory whose validity in the parameter regime $|\mu|\sim1/4$ is correct
only in a perturbative sense such as neglecting a higher frequency term $\cos(8\sqrt{\pi})$.
Thus, authors of \cite{Furukawa} also carried out an independent iTEBD calculation of the
original model (1), confirming only one dimer to N\'{e}el transition at $\Delta \sim 0.72$ and one
N\'{e}el to dimer transition at $\Delta \sim 0.93$. It was reported that iTEBD experienced a 
poor convergence for $0.95\leq\Delta<1$ and $0.25<|\mu|<0.35$. 

The purpose of this letter is to study the controversial parameter regime in the above by EPT
(entanglement perturbation theory), a recent many body method \cite{Chung1,Chung2,Chung3,Chung4}.
Our finding in this paper is summarized as follows: (1) EPT gives a converging, and hence exact phase diagram, including exact order parameters of N\'{e}el and dimer phases for the entire $0\leq\Delta<1$ along the Lifshitz line, see Fig.4 below.
The dimer phase and N\'{e}el phase alternate only once, and for $\Delta \geq 0.93$, only the dimer phase exists. This means two points. 
(2) First, the perturbative SG analysis when applied to the deeply frustrative regime $|\mu| \sim 1/4$, is missing some physics, 
and (3) second, iTEBD, even with a large entanglement $\chi \sim 300$, is not as accurate as EPT with $\chi \sim 80$. 
This conclusion is not a surprise. Indeed, TEBD operates only on a few local spins, and therefore the deterioration of accuracy 
due to loss of information arising from successive Hilbert space truncation, typical of real-space RG, is expected worse than DMRG.

The EPT algorithms EPT-g1 and EPT-g2 both solve for ground state properties of systems 
with translational symmetry (a recent development of EPT handling inhomogeneity will be reported elsewhere). EPT-g1 is especially suitable for 
infinite systems and is our starting point. It was previously applied to the $J_2=0$ case
 \cite{Wang,Wang2}, and its superb accuracy is demonstrated in the first confirmation of
a prediction made by the conformal field theory \cite{Affleck,Fisher} on a long-range 
spin-spin correlation which can only be seen for thousands of lattice separation. 
It also confirmed yet another field theory prediction, by bosonaization, for the 4-spin correlation
functions with better precision than DMRG \cite{Hikihara}. 

The central idea of EPT-g1 is as follows. First is the system wave function in the matrix product state 
(MPS) \cite{Chung1,MPS1,MPS2,MPS3,MPS4,MPS5} representation 
\begin{align}
\label{eq:1DEPT}
|\psi\rangle
=\sum_{\{\sigma\}}{\rm Tr}\left[\xi_1^{\sigma_1}\cdot
\xi_2^{\sigma_2}\cdots\xi_N^{\sigma_N}\right]|\sigma_1\rangle\otimes|\sigma_2\rangle
\otimes\cdots\otimes|\sigma_N\rangle 
\end{align} 
where $\xi_i^{\sigma_i}$ is the local wave function matrix at site $i$, 
and $N$ is the number of sites. $\sigma_{i}(=\uparrow,\downarrow)$ refers to local states. 
The dimension $p$ of the square matrices $\xi_i^{\sigma_i}$, the entanglement, 
controls the accuracy of the wave function $|\psi\rangle$. In principle $|\psi\rangle$ 
with optimized $\xi_i^{\sigma_i}$ becomes 
exact with $p\to\infty$. The matrices $\xi_i^{\sigma_i}$ are optimized to maximize the 
variational energy: 
\begin{align}
\label{eq:maineqn}
\frac{\partial}{\partial \xi_i} \frac{\langle \psi\left(\xi_1,\xi_2,\cdots,\xi_N\right)\mid 
e^{-\beta H} \mid \psi\left(\xi_1,\xi_2,\cdots,\xi_N\right)\rangle}
{\langle \psi\left(\xi_1,\xi_2,\cdots,\xi_N\right)\mid 
\psi\left(\xi_1,\xi_2,\cdots,\xi_N\right)\rangle}=0 
\end{align}
where $\beta \rightarrow 0$ \cite{Chung2}. We then arrive at a 
generalized eigenvalue equation 
\begin{align}
\label{eq:eptgeigen}
& X_i\left(\xi_1,\cdots,\xi_{i-1},\xi_{i+1},\cdots,\xi_N\right)\xi_i \notag\\
= & \epsilon Y_i\left(\xi_1,\cdots,\xi_{i-1},\xi_{i+1},\cdots,\xi_N\right)\xi_i  \quad (i=1,2\cdots,N), 
\end{align}
where $X_i$ and $Y_i$ are symmetric matrices depending on $\{\xi_i\}$, and $\epsilon$ 
is the eigenvalue. The eigenfunction $\{\xi_i\}$ with the largest $\epsilon$ describes 
the ground state wave function $|\psi\rangle$.

\begin{figure}
\begin{center}
\includegraphics[width=18pc]{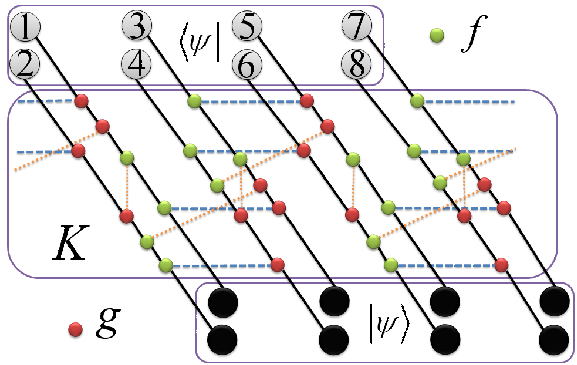}
\caption{\label{fig:K_zigzag}Schematic figure of the MPO of the density matrix for a zigzag chain. 
Short dashed lines (orange) indicate NN interactions and long dashed lines (blue) NNN interactions. 
Solid lines refer to the inner prodoct.}
\end{center}
\end{figure}

\begin{figure}
\begin{center}
\includegraphics[height=1.7in, width=2.5in]{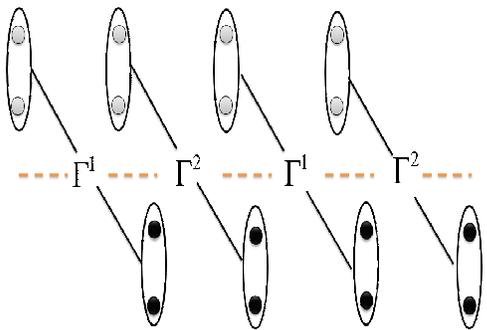}
\caption{\label{fig:K_final}Simplified schematic figure of the MPO of the density matrix for a zigzag chain. 
The two spins enclosed by an ellipse form an effective site. The dashed lines (orange) indicate the direct product and solid lines the inner product.}
\end{center}
\end{figure}

The density matrix $K \equiv e^{-\beta H}$ can also be written in matrix product form, the 
matrix product operator (MPO) \cite{Chung1,Wang2}. A generalization to quasi-1D naturally 
inherits this MPO of density matrix along with MPS for the wave function. But now, cf. Fig.2,
 MPS is defined on a 2-spin composite (grey and black vertical two circles) 
and MPO is in a complex shell by shell (inner product direction)
and layer by layer structure (vertical direction). 
We first deform a zigzag chain into two chains shown in Fig.\ref{fig:deformation}. 
The Hamiltonian is rewritten as a summation of local bond Hamiltonians, 
\begin{align}
\label{EQ:SUMHBOND}
H=\sum_{bond}H_{bond}
\end{align}
There are four types of bonds. Bond $\left\{1,2\right\}$, $\left\{3,4\right\}$ etc form the 
first series. $\left\{2,3\right\}$, $\left\{4,5\right\}$ etc belong to the second. 
These two series account for NN interactions. The third is 
$\left\{1,3\right\}$,$\left\{2,4\right\}$,$\left\{5,7\right\}$,$\left\{6,8\right\}$ etc 
and the forth is $\left\{3,5\right\}$,$\left\{4,6\right\}$,$\left\{7,9\right\}$,$\left\{8,10\right\}$ etc. 
They account for NNN interactions. The way to group is not unique, but it does not affect the result. 
It is immediately seen that $\left[H_{bond_i},H_{bond_j}\right]=0$ if they are in the same series. This is the criteria to group the bonds. Small $\beta$ safely separates series in the density matrix as follows
\begin{align}
\label{eptg1speratesublattice}
e^{-\beta H} & =\prod_{bond}{e^{-\beta H_{bond}}}+{\mathcal{O}}\left(\beta^2\right)\notag\\
& \approx \cdots e^{-\beta \sum_{1-st~series}{H_{bond}}}e^{-\beta\sum_{2-nd~series}{H_{bond}}}\cdots
\end{align}
And for each series, we have 
\begin{align}
\label{eq:series1}
e^{-\beta \sum_{series}{H_{bond}}}=\cdots f_{\alpha}\otimes & g_{\alpha} \otimes f_{\gamma} \otimes g_{\gamma}\cdots +{\mathcal{O}}\left(\beta^2\right)
\end{align}
where repeating indexes follow Einstein summation convention. $f$, for instance for the XXX model, 
takes four $2\times 2$ matrices: $1,\sqrt{\beta}S^x,\sqrt{\beta}S^y,\sqrt{\beta}S^z$ on a site. 
And $g$ likewise on the other site of a bond. We use $\it{shell}$ to denote 
a series in (\ref{eq:series1}). There are four shells in total. They are coupled in the direction of in and out of the paper 
in Fig.\ref{fig:K_zigzag}, by the inner product of the operators on the same site. 
Also note that MPO contains two $\it{layers}$, composed of two vertically aligned spins, one in the 
top chain and the other in the bottom chain. 
We finally have an explicit expression of the whole density matrix as follows
\begin{align}
	\label{eq:K1}
	K\equiv \cdots \Gamma^1_{\alpha\beta\gamma,\alpha'\beta'\gamma'}\otimes \Gamma^2_{\alpha'\beta'\gamma',\alpha''\beta''\gamma''}
\end{align}
where
\begin{align}
	\label{eq:gamma1}
	& \Gamma^1_{\alpha\beta\gamma,\alpha'\beta'\gamma'}\notag\\
 =& f_{\alpha'}\left(\mu J_1\right)f_i\left(J_1\right)g_{\beta}\left( J_1\right)g_{\alpha}\left(\mu J_1\right)\notag\\
 & \hspace{.9in}  \otimes\\
& f_{\gamma'}\left(\mu J_1\right)f_{\beta'}\left(J_1\right)g_{i}\left( J_1\right)g_{\gamma}\left(\mu J_1\right)\notag
\end{align}
and $\Gamma^2$ likewise. $\alpha$,$\beta$ and $\gamma$ etc run from 1 to 4. They are the indexes 
involved in the direct product. If they are absorbed into an effective index, 
both $\Gamma^1$ and $\Gamma^2$ become 4-leg tensors. The two effective legs run from 
1 to $4^3=64$. And the other two 
legs run from 1 to 4, responsible for the inner product with a new local wave function $\xi$
now defined on a 2-spin composite, denoted by an ellipse in Fig.\ref{fig:K_final}. 
It is also noted that, reflecting the $\Gamma^{1,2}$ repetition structure, the local wave functions
are also of two species $\xi^{1,2}$.  The generalized eiegnvalue problem (4) is solved iteratively 
for the local wave functions, and after convergence, energy etc can be calculated as before 
\cite{Chung2,Wang2}. 

The calculation on zigzag chains is more expensive compared with a single spin chain 
because one now has to 
deal with the larger number of local states and larger indexes in MPO. Let us discuss the major 
time consumption. There are two. The first are generalized eigenvalue equations whose size 
is $np^2$ with $n$ being the number of effective local states. The second is an 
eigenvalue decomposition of a transfer matrix in the horizontal direction in Fig.3, 
a basis unit involving $\Gamma^1$ and $\Gamma^2$ and in total 4 2-spin composites 
connected to $\Gamma^{1,2}$. 
Its size is $mp^2$ where $m$ is the size of the 3 bonds in the direct product between 
$\Gamma^1$ and $\Gamma^2$. In contrast to $n=2$ and $m=4$ in a spin chain, they are $4$ and $64$ now. 
It is seen that the eigenvalue decomposition of $mp^2$ x $mp^2$ transfer matrix becomes especially large. 

Fortunately, there is a simple way to overcome this difficulty. 
The key point is in the original idea of EPT-g1, namely convert the Hamiltonian eigenvalue problem to
that for the density matrix $K \equiv e^{-\beta H}$, noting $e^{-\beta H} \rightarrow 1-\beta H$,
leading to an MPO representation of the density matrix. A simple example is the product of two 
local density matrixes, $(1-\beta H_{ij})\otimes (1-\beta H_{kl}) \sim 1-\beta H_{ij}-\beta H_{kl}+
\beta^2H_{ij}\otimes H_{kl}$
, where the last term is unimportant in the $\beta \rightarrow 0$ limit. $\beta=10^{-6} \sim 10^{-7}$
is used in this work like before \cite{Chung2}.  Likewise, 
$\Gamma^{1,2}_{\alpha\beta\gamma,\alpha'\beta'\gamma'}$ should contain such higher order 
terms in $\beta$ which we can consistently discard. A rule to do so is in the fact that,
in $f_{\alpha}$ and $g_{\alpha}$ matrixes, $\alpha=1$ is identity matrix while other three terms are
all order $\sqrt{\beta}$. We should keep in mind that whenever there is a term of order 
$\sqrt{\beta}$, there should be a companion order $\sqrt{\beta}$ term at the other end of the bond.
This amounts to keep, among 64 index combinations in $\alpha\beta\gamma$, those terms with only one of
$\alpha\beta\gamma$ not being $1$, namely, $111,112,121,211,113,131,311$ and $114,141,411$, there are total $10$.
 
To test the algorithm, we have calculated the dimerization 
$d\equiv \left\langle \vec{S}_i\left(\vec{S}_{i-1}-\vec{S}_{i+1}\right)\right\rangle$ for 
$\mu=0.5871$ and $\Delta=0$ where $d$ is maximum when both $Js$ are positive \cite{White}. 
The EPT result is $0.7906137$ when $p=35$ vs. $0.7906135$ by DMRG \cite{White}. 
By comparing with their results for $\mu=2.5$ and $\Delta=0$, the most difficult case studied 
in \cite{White} and probably the same for EPT, it is found that EPT converges faster  
in entanglement. Since $p$-equivalent $\chi=200\sim 300$ is used in iTEBD \cite{Furukawa1}, 
the entanglement needed by EPT to achieve the comparable accuracy is much smaller than iTEBD. 

\begin{figure}
\begin{center}
\includegraphics[height=2.2in, width=3.3in]{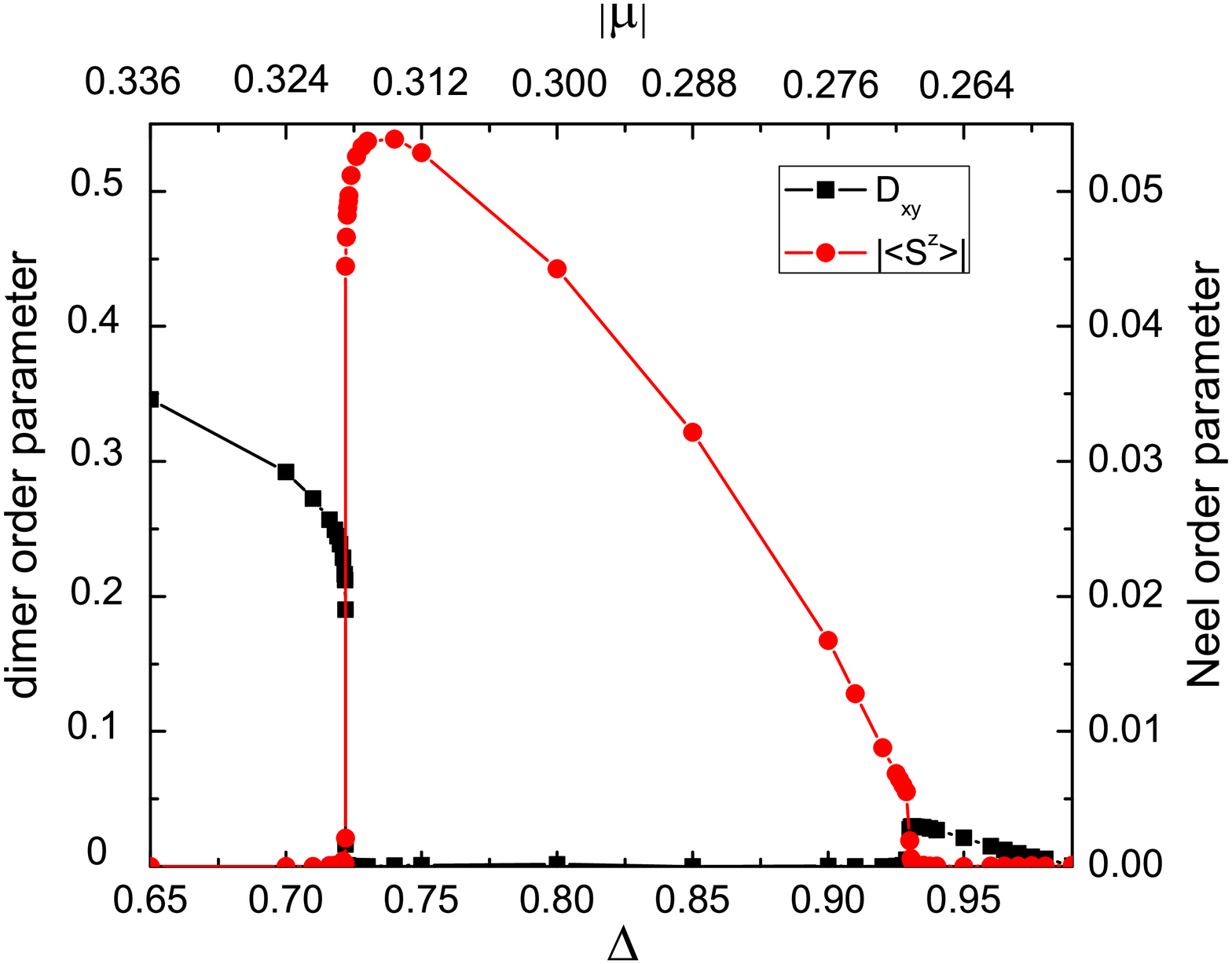}
\caption{\label{fig:phase_diagram} The phase diagram along the line connecting $\left(\left|\mu\right|,\Delta\right)=\left(0.336,0.65\right)$ and $\left(0.264,0.95\right)$, which approximately traces the Lifshitz line. }
\end{center}
\end{figure}

The main result of this letter is the phase diagram shown in Fig.\ref{fig:phase_diagram} 
where $\Delta$ approaches the isotropic point $1$ from below along the Lifshitz line.
The N\'{e}el order parameter is $\left\langle S_i^z\right\rangle$ and the dimer 
order parameter $D_{xy}$ is defined by
\begin{align}
	D_{xy}\equiv \left(S_{i-1}^x S_{i}^x +S_{i-1}^y S_{i}^y\right)- \left(S_{i}^x S_{i+1}^x +S_{i}^y S_{i+1}^y\right)
\end{align}

The main difference between our phase diagram and that of \cite{Furukawa} is that our calculation converged
solidly for all $\Delta$ up to $1$, whereas iTEBD failed for $0.95<\Delta<1$. It also shows no more 
alternation of the N\'{e}el and dimer orders for $\Delta > 0.93$, disproving the prediction
of \cite{Furukawa} that there should be an infinite number of such transitions. The failed prediction, we
argue, is most likely due to the failure of perturbative analysis of the Sine-Gordon theory deep 
in the frustrated regime $|\mu| \sim 1/4$. 
Technically, the Lanczos diagonalization not only fails to converge due to highly degenerate nature for $0.995\leq\Delta\leq 1$ 
as pointed out in \cite{Furukawa}, but also, we argue, yields not quite accurate energy levels for $\Delta>0.95$ 
so that the level spectroscopy analysis concludes incorrectly for this parameter regime. 
As for the nature of the N\'{e}el to dimer transition, 
our calculation, Fig.\ref{fig:phase_diagram}, does not agree with \cite{Furukawa} that it is first order. 
Rather it appears second order.

\begin{figure}
\begin{center}
\includegraphics[height=2.3in, width=2.8in]{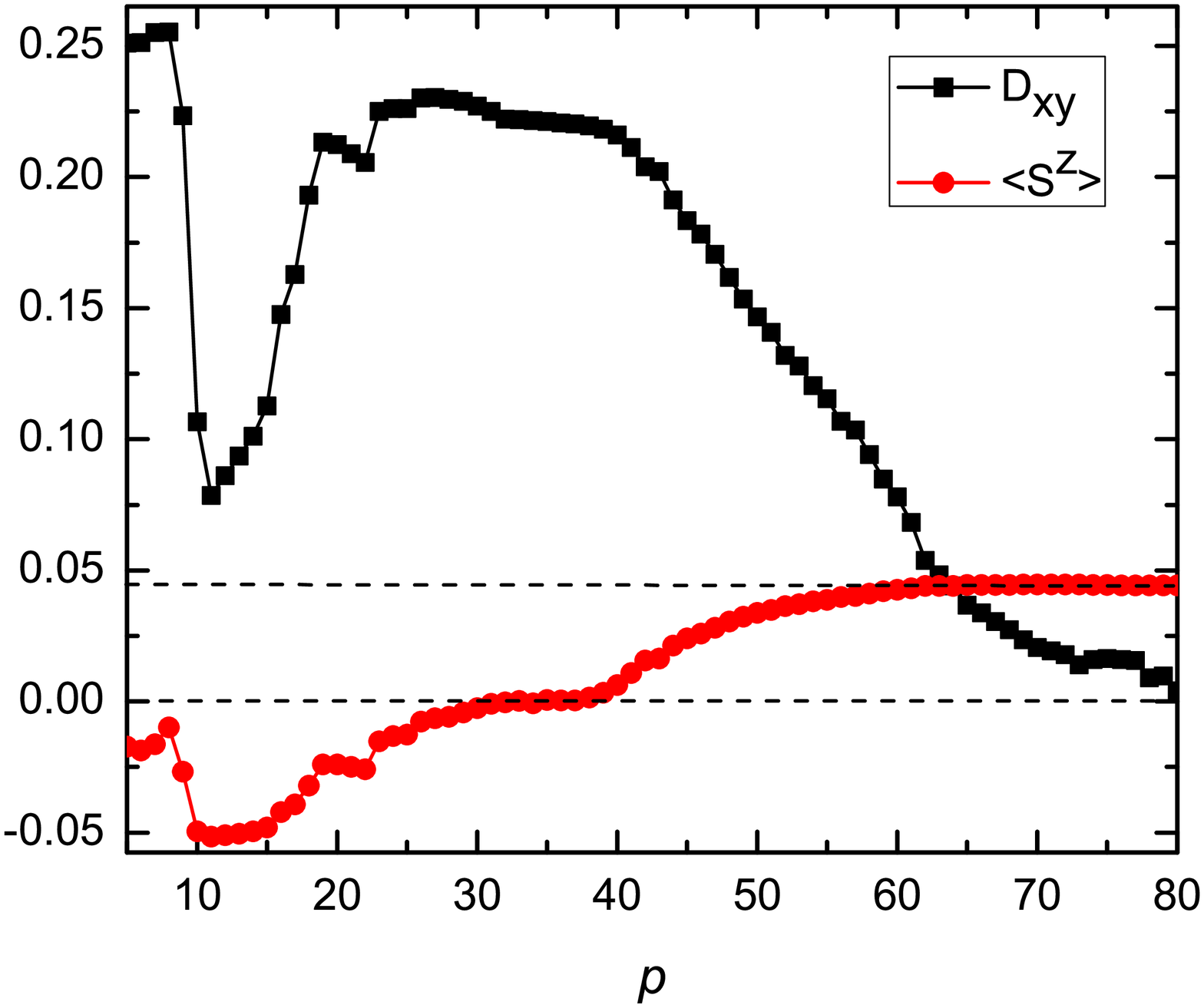}
\caption{\label{fig:phaseboundary} Order parameters vs entanglement $p$ at $\Delta=0.722$, 
one of the phase boundaries. N\'{e}el order wins the competition only at a large $p$.}
\end{center}
\end{figure}  

While our calculation converged solidly, there is a clear evidence of a hard competition
between the N\'{e}el and dimer orders near the transition point. 
In Fig.\ref{fig:phaseboundary}, we have plotted the N\'{e}el 
and dimer order parameters as a function of entanglement for $\Delta=0.722$.
Dimer order dies out only at relatively large entanglement.
The same happens at the other phase boundary $\Delta=0.930$ with even more subtle 
competition between the two orders since their magnitudes are both small.

To summarize, we have developed EPT for quasi-1D strongly correlated quantum systems. 
There are two immediate applications of interest. 
First is the 1D Hubbard model with long-range hopping, long range Coulomb interaction,
orbital degeneracy and Hund-coupling in the pursuit of itinerant 
ferromagnetism and triplet superconductivity.
Second is a spin tube, namely the effective sites now contain the third spin, taking into account third
nearest neighbor exchange interactions. Works are currently under way, and will be reported elsewhere.

Note added -- After the completion of the paper, we have heard from the authors of \cite{Furukawa} that
the Lifshitz line along which we and they have done the calculations is actually not precise, and that their prediction of the N\'eel phase could still be observed close to the ferromagnetic point, $\Delta= 1.0$. We have explored $\Delta = 0.98$ and $J_2/|J_1| = 0.2552$ (newly suggested point) to 0.2568 (our and their previous calculation point). We observed a TLL ($J_2/|J_1|=0.2552$) to dimer order transition of the second order type which is indistinguishable from the 
BKT (Berezinskii-Kosterlitz-Thouless) type \cite{bere,kt} due
to an essential singularity\cite{chung8}, the transition point being 0.2560. 
We have also calculated the chiral order parameter $\mathbf{S_1}\textrm{x}\mathbf{S_2}$
in this parameter region to be strictly zero. In the meantime, we have observed a finite 
chiral order parameter for the point $\Delta=0.9$ and $J_2/|J_1|=0.35$ which is 
deeply in the chiral phase. We thus conclude that our main claim in the paper is unchanged. 

The numerical calculation utilized the RIKEN Integrated Cluster of Clusters at Advanced 
Center for Computing and Communication, RIKEN. We thank the authors of \cite{Furukawa}
for useful discussions.

\bibliography{zigzagchain}

\end{document}